\documentclass[showpacs,aps,twocolumn]{revtex4}

\usepackage{bm}
\usepackage{amsmath}
\usepackage{graphicx}
\usepackage{subfigure}
\usepackage[usenames,dvipsnames]{color}
\definecolor{darkblue}{RGB}{0,0,196}
\usepackage[colorlinks=true,linkcolor=darkblue,citecolor=darkblue,urlcolor=darkblue]{hyperref}

\usepackage{setspace}
\usepackage{footmisc}
\usepackage[makeroom]{cancel}
\usepackage{comment}
\usepackage{hyperref}

\usepackage{color,soul}

\def\be{\begin{equation}}
\def\ee{\end{equation}}
\def\ba{\begin{eqnarray}}
\def\ea{\end{eqnarray}}

\usepackage{graphicx}
\usepackage{amsmath,bbm}
\usepackage{amssymb,bm}

\begin{document}

\title{Wiedemann-Franz Law For Hot QCD Matter in Color String Percolation Scenario}
\author{Pragati Sahoo}
\email{pspragatisahoo@gmail.com}
\author{Raghunath~Sahoo}
\email{Raghunath.Sahoo@cern.ch (Corresponding Author)}
\affiliation{Discipline of Physics, School of Basic Sciences, Indian Institute of Technology Indore, Indore- 453552, INDIA}
\author{Swatantra~Kumar~Tiwari}
\email{sktiwari4bhu@gmail.com}
\affiliation{Department of Applied Science and Humanities, Muzaffarpur Institute of Technology, Muzaffarpur- 842003, Bihar}

\begin{abstract}
\noindent
Transport coefficients serve as important probes in characterizing the QCD matter created in high-energy heavy-ion collisions. Thermal and electrical conductivities as transport coefficients have got special significance in studying the time evolution of the created matter. We have adopted color string percolation approach for the estimation of thermal conductivity ($\kappa$), electrical conductivity ($\sigma_{el}$) and their ratio, which is popularly known as Wiedemann-Franz law in condensed matter physics. The ratio $\kappa/\sigma_{el}T$, which is also known as Lorenz number ($\mathbb{L}$) is studied as a function of temperature and is compared with various theoretical calculations. We observe that the thermal conductivity for hot QCD medium is almost temperature independent in the present formalism and matches with the results obtained in ideal equation of state (EOS) for quark-gluon plasma with fixed coupling constant ($\alpha_s$). The obtained Lorenz number is compared with the Stefan-Boltzmann limit for an ideal gas. We observe that a hot QCD medium with color degrees of freedom behaves like a free electron gas.

\end{abstract}


\date{\today}

\maketitle 
\section{Introduction}
\label{intro}

Ultrarelativistic heavy-ion collision programmes such as Relativistic Heavy-Ion Collider (RHIC) at BNL and Large Hadron Collider (LHC) at CERN possibly create a deconfined state of quarks and gluons called quark-gluon plasma (QGP). At these experimental facilities, an ample amount of data have been collected, which help us to understand the thermodynamic and transport properties of the created matter. Analysis of the experimental observables such as transverse momentum spectra and collective flow of charged hadrons or electromagnetic probes necessitates the inclusion of transport parameters in their estimation. This motivates to study the transport properties of the hot QCD matter produced in heavy-ion collisions. The first experimental proof of the transport processes in the created medium is given by explaining the charged hadrons elliptic flow measured at RHIC~\cite{STAR} using dissipative hydrodynamics~\cite{Luzum}. Recently, ALICE results~\cite{ALICE1, ALICE2, ALICE3, ALICE4, ALICE5, ALICE-JHEP} have reconfirmed the relevance of these transport processes. Currently, the transport coefficients under intense investigations are shear viscosity ($\eta$), bulk viscosity ($\zeta$), electrical conductivity ($\sigma_{\rm el}$) and thermal conductivity ($\kappa$).

The space-time evolution of the system formed in ultra-relativistic high energy collisions can be quantified by the energy-momentum dissipation. The velocity gradient between the adjacent layers of the system resulting distortion in momentum distribution of the system quanta gives rise to viscous forces. Like momentum distortion, thermal dissipation also happens due to temperature gradient in the system. This can be described in terms of thermal conductivity for a system with conserved baryon current density. A strong electromagnetic field is generated in the early stages of non-central heavy-ion collisions, which is of the order of $m_{\pi}^2$. To quantify the impact of the fields on electromagnetically charged QCD medium produced, the electrical conductivity plays an important role. It gives a measure of the electric current being induced in the response of the early stage electric field~\cite{Mitra:2017sjo}. The ratios between thermal and electrical conductivity can reflect the competition between momentum transport, heat transport in the medium, respectively, leading to the verification of Wiedemann-Franz Law for a QCD medium. The electrical conductivity plays an important role in quantifying the impact of the fields on electromagnetically charged medium. It measures the electric current induced in the response of the early stage electric field. However, the thermal conductivity has a crucial role in the hydrodynamic evolution of the medium created specially at FAIR energies and in the low energy runs at RHIC, where baryon chemical potential is significant. In Ref.~\cite{Kapusta}, it is shown that the thermal conductivity diverges at the critical point and also used to study the impact of hydrodynamic fluctuations on experimental observables. Recently, various transport coefficients such as electrical conductivity and thermal conductivity for QGP are studied using the quasi-particle model. In these works, the Chapman-Enskog technique is employed from kinetic theory of many particle system with a collision term that includes the binary collisions of quarks/antiquarks and gluons~\cite{Mitra:2017sjo,Mitra:2018akk}. 

Color String Percolation Model which is inspired by QCD~\cite{Armesto:1996kt,Nardi:1998qb,Braun:1999hv,Braun:1997ch,Braun:2000hd}, can be used as an alternative approach to Color Glass Condensate (CGC). The color flux tubes are stretched between the colliding partons in terms of the color field in the color string percolation scenario. ${\it q\bar q}$ pair is produced from strings in a similar way as in the Schwinger mechanism of pair creation in a constant electric field covering all the space~\cite{Phyreport}. When the energy and the number of nucleons of participating nuclei increase, the number of strings grows. Color strings may be viewed as small discs in the transverse space filled with the color field created by colliding quarks and gluons. The number of strings grows as collision energy and size of the colliding nuclei increase and starts overlapping to form clusters. When a critical string density is reached, a macroscopic cluster appears that marks the percolation phase transition which spans the transverse nuclear interaction area. 2-dimensional percolation is a non-thermal second order phase transition. In CSPM, the Schwinger barrier penetration mechanism for particle production, the fluctuations in the associated string tension and the quantum fluctuations of the color fields make it possible to define a thermodynamical equilibrium temperature. Consequently, the particle spectrum is produced with a thermal distribution. When the initial density of interacting colored strings ($\xi$) exceeds the 2D percolation threshold ($\xi_c$) i.e. $\xi > \xi_c $, a macroscopic cluster appears, which defines the onset of color deconfinement. The critical density of percolation is related to the effective critical temperature and thus percolation may be a possible way to achieve deconfinement in heavy-ion collisions \cite{PLB642} and in high multiplicity pp collisions~\cite{Gutay:2015cba,Hirsch:2018pqm}. It is observed that, CSPM can be successfully used to describe the initial stages in ultrarelativistic high energy collisions \cite{Phyreport}. Recently, we have studied the temperature dependence of electrical conductivity and shear viscosity to entropy density for strongly interacting matter using CSPM~\cite{Sahoo:2018dxn}. We have also studied the collision energies, collision centrality and colliding system size dependence of various thermodynamical and transport properties for the strongly interacting matter in a color string percolation approach~\cite{Sahoo:2017umy,Sahoo:2018dcz}.  In our earlier work \cite{Sahoo:2018dxn}, we had an extensive study of electrical conductivity of hot QCD matter using the CSPM approach. In the present report, we extend the work to include the thermal conductivity and study the famous Wiedemann-Franz law of condensed matter physics for a system with color degrees of freedom possibly created in heavy-ion collisions at the relativistic energies or in high-multiplicity proton-proton collisions at the LHC energies. 

The manuscript is organized as follows: Section~\ref{formulation} encompasses the derivation of electrical and thermal conductivities in color string percolation model (CSPM). The obtained results using the formulations have been discussed in Section~\ref{RD}. Finally, we present the summary of the work with possible outlook  in Section~\ref{summary}. 

\section{Formulation}
\label{formulation}
In this section, we present the derivation of the electrical and thermal conductivities using CSPM for hot QCD matter and study the temperature dependence of the Lorenz number.

\subsection{Electrical Conductivity} 
\label{elec}

First, we formulate the electrical conductivity of strongly interacting matter using the color string percolation approach. To start with we need to revisit the basic formalism of CSPM. A parameterisation of pp collisions at $\sqrt{s}$ = 200 GeV for central Au+Au collisions is used to compute the percolation density parameter, $\xi$ by using the $p_T$ distribution. And the parameterisation is given as~\cite{Sahoo:2018dxn,Sahoo:2017umy,Sahoo:2018dcz},


\begin{eqnarray}
\frac{dN_{\rm ch}}{dp_{\rm T}^{2}} = \frac{a}{(p_{\rm 0}+{p_{\rm T}})^{\alpha}},
\end{eqnarray}

where, a is the normalisation factor and  $p_{0}$, $\alpha$ are fitting parameters given as, $p_{0}$ = 1.982 and $\alpha$ = 12.877~\cite{Phyreport}. 
 Due to the low string overlap probability in pp collisions there is less probability for percolation to happen, so the fit parameters extracted by fitting $p_T$ distribution of pp collisions at  $\sqrt{s}$ = 200 GeV are used to evaluate the interactions of the strings in Au+Au collisions. The modified parameterisation is given as,


\begin{eqnarray}
 p_{\rm 0}\rightarrow p_{\rm 0}\left(\frac{\langle  nS_{\rm 1}/S_{\rm n}\rangle_{\rm Au+Au}}{\langle nS_{\rm 1}/S_{\rm n}\rangle_{\rm pp}}\right)^{1/4}.
 \label{po}
\end{eqnarray}

Here, $S_{\rm n}$ represents the area occupied by $n$ overlapping strings. Now,  

\begin{eqnarray}
 \langle \frac{nS_{\rm 1}}{S_{\rm n}} \rangle = \frac{1}{F^{\rm 2}(\xi)},
 \label{p1}
\end{eqnarray}

where, $F(\xi)$ is the color suppression factor, which is given as,

\begin{eqnarray}
 F(\xi) = \sqrt \frac{1-e^{-\xi}}{\xi}.
 \label{p1}
\end{eqnarray}

To calculate the electrical conductivity of strongly interacting matter, which is one of the most important transport properties of QCD matter, we proceed as follows. The mean free path $(\lambda_{\rm mfp})$, which denotes the relaxation of the system far from equilibrium can be written in terms of number density ($n$) of an ideal gas of quarks and gluons and the transport cross-section ($\sigma_{\rm tr}$). i.e.

\begin{eqnarray}
\lambda_{\rm mfp} = \frac{1}{n\sigma_{\rm tr}}.
\label{eq5}
\end{eqnarray}

In CSPM the number density is given by the effective number of sources per unit volume:

\begin{eqnarray}
n = \frac{N_{\rm sources}}{S_{\rm n}L}.
\label{el}
\end{eqnarray}

Here, $L$ is the longitudinal extension of the string $\sim $1 fm. The area occupied by the strings is given by the relation $(1 - e^{-\xi})S_{\rm n}$. Thus, the effective number of sources is given by the total area occupied by the strings divided by the area of an effective string, $S_1F({\xi})$ as shown below,
 
\begin{eqnarray}
N_{\rm sources} = \frac{(1 - e^{-\xi})S_{\rm n}}{S_{\rm 1}F({\xi})},
\label{el}
\end{eqnarray}

In general, $N_{\rm sources}$ is smaller than the number of single strings. In the limit, $\xi = 0$, $N_{\rm sources}$ equals to the number of strings $N_{\rm s}$. So,

\begin{eqnarray}
n = \frac{(1 - e^{-\xi})}{S_{\rm 1}F({\xi})L}.
\label{eq8}
\end{eqnarray}

Now, using eqs.~\ref{eq5} and~\ref{eq8}, the $\lambda_{\rm mfp}$ can be expressed in terms of percolation density parameter,

\begin{eqnarray}
\lambda_{\rm mfp} = \frac{L}{(1 - e^{-\xi})},
\label{mfp}
\end{eqnarray}

where $\sigma_{\rm tr}$, the transverse area of the effective strings equals to $S_1F(\xi)$.  

Now we use Anderson-Witting model to derive the formula for electrical conductivity in which the Boltzmann transport equation is given as~\cite{Anderson},


\begin{eqnarray}
p^\mu\partial_\mu f_k + qF^{\alpha\beta}p_{\beta}\frac{\partial f_k}{\partial p^{\alpha}} = \frac{-p^\mu u_{\mu}}{\tau}(f_k - f_{eq,k}),
\label{bte}
\end{eqnarray}

where $f_k = f(x,\overrightarrow{p},t)$ is the full distribution function and $f_{eq,k}$ is the equilibrium distribution function of $\rm k^{th}$ species. $\tau$ is the mean time between collisions and $u_\mu$ denotes the fluid four velocity in the local rest frame. Eq.~\ref{bte} provides the calculation of the quark distribution after applying the electric field. The gluon distribution function remains thermal and unaffected by electric field. The assumptions that, there are as many quarks (charge $q$) as anti-quarks (charge -$q$) and uncharged gluons in the system is considered. $F^{\alpha\beta}$ is the electromagnetic field strength tensor, which in terms of electric field and the magnetic flux tensor is given as~\cite{Greif:2014oia},


\begin{eqnarray}
F^{\mu\nu} = u^{\nu} E^{\mu} - u^{\mu} E^{\nu} - B^{\mu\nu}.
\label{field}
\end{eqnarray}
The magnetic field is set to zero, $B^{\mu\nu} = 0$ in the calculations, as we are interested in the study of the effect of electric field on the system.  The electric current density of the $\rm k^{th}$ species in the $x$-direction is given as,

\begin{eqnarray}
j^x_k = q_k\int \frac{d^3p p^x}{(2\pi)^3 p^0}f_k = g_k\tau\frac{8}{3}\frac{\pi q_{k}^2 T^2}{(2\pi)^3}E^x.
\label{Current}
\end{eqnarray}
According to Ohm's law, $j^x_k = \sigma_{el}E^x$. Using eq.~\ref{Current} and relation $n_k = g_k T^3/\pi^2$, electrical conductivity in the assumption of very small electric field and no cross effects between heat and electrical conductivity in the relaxation time approximation is given by,

\begin{eqnarray}
\sigma_{\rm el} = \frac{1}{3T} \sum_{\rm k=1}^{M}q_{\rm k}^2 n_{\rm k} \lambda_{\rm mfp}.
\label{el_con}
\end{eqnarray}

Putting the expression of $\lambda_{\rm mfp}$ in eq.~\ref{el_con} and only considering the density of up quark $(u)$ and its antiquark $(\bar{u})$ in the calculation, we get the expression for $\sigma_{\rm el}$ as, 

\begin{eqnarray}
\sigma_{\rm el} = \frac{1}{3T}\frac{4}{9}e^{2}n_{\rm q}(T) \frac{L}{(1 - e^{-\xi})}.
\label{el_c}
\end{eqnarray}

Here, the pre-factor 4/9 reflects the fractional quark charge squared $( \sum_{\rm f} q_{\rm f}^{2}; \ q_{\rm f} = \frac{2}{3})$ and $n_{\rm q}$ denotes the total density of quarks or antiquarks. Here, $e^2$ in the natural unit is taken as $4\pi\alpha$, where $\alpha$ = 1/137.

\subsection{Thermal Conductivity} 

Thermal conductivity ($\kappa$) is another interesting observable, which describes the heat flow in interacting systems~\cite{Israel:1979wp,groot} and gained a recent interest in ultrarelativistic heavy-ion collisions~\cite{Denicol:2012cn,Greif:2013bb}. $\kappa$ in CSPM is derived as follows:

The heat current density, $j_{Q}$, i.e. the amount of thermal energy crossing an unit area per unit time is proportional to the temperature gradient,

\begin{eqnarray}
 J_{Q} = -\kappa \frac{dT}{dx},
 \label{e1}
\end{eqnarray}

where $\kappa$ is the thermal conductivity. Hot and dense matter also called fireball created in heavy-ion collisions gives rise to numerous partons, which involve in the process of heat conduction in the fireball. Heat conduction happens both by quarks and gluons in the form of thermal energy.

From kinetic theory of gases, the thermal conductivity, $\kappa$ is evaluated by using the formula, $\kappa = (1/3) C_{V} v_{F}^2 \tau $~\cite{link,Heiselberg:1993cr}. Here $C_{V}$ is the specific heat per unit volume, $v_F$ is the Fermi velocity and $\tau$ is the mean free time of partons. Now, $\kappa$ using the expression of specific heat~\cite{link1} is written as,

\begin{eqnarray}
 \kappa = \frac{1}{3} \big(\frac{\pi^2 n k_{B}T}{2T_{F}})v_{F}^2\tau.
 \label{e3}
\end{eqnarray}

Here the Fermi temperature ($T_F$) is defined as, $T_F = E_F/k_B$ and Fermi energy, $E_F$ can be expressed as $\frac{1}{2}mv_F^2$. Now the above equation can be expressed for a system of partons as,

\begin{eqnarray}
 \kappa =  \frac{\pi^2 n k_{B}^{2}\lambda_{\rm mfp} T}{3m}
 \label{e4}
\end{eqnarray}

In natural unit, $k_{B}$ = 1 and $m$ can be written as, $m$ = 3$T$ , thermal conductivity for strongly interacting matter then becomes,

\begin{eqnarray}
 \kappa =  \frac{\sum_{k}\pi^2 n_{k}\lambda_{\rm mfp}}{9}
 \label{e4}
\end{eqnarray}

In the context of CSPM, the above equation is reduced by using the expression of $\lambda_{\rm mfp}$ as,

\begin{eqnarray}
 \kappa =  \frac{\pi^2}{9}\frac{ \sum_{k = partons}n_{k}L}{(1-e^{-\xi})}
 \label{e4}
\end{eqnarray}

The relation between electrical conductivity $\sigma_{el}$, and thermal conductivity $\kappa$, for any substance can be understood in terms of the Wiedemann-Franz law. The basic mathematical statement of the law is,

\begin{eqnarray}
 \kappa/\sigma_{el}T =  \frac{\pi^2}{3e^2}\equiv \mathbb{L}
 \label{e4}
\end{eqnarray}

The Lorentz number, $\mathbb{L}$ is a constant which quantifies the system as good electrical as well as thermal conductors. To understand the quantum aspects of its liquidity of the QCD system Wiedemann-Franz law becomes relevant.

\begin{figure}
\includegraphics[height=20em]{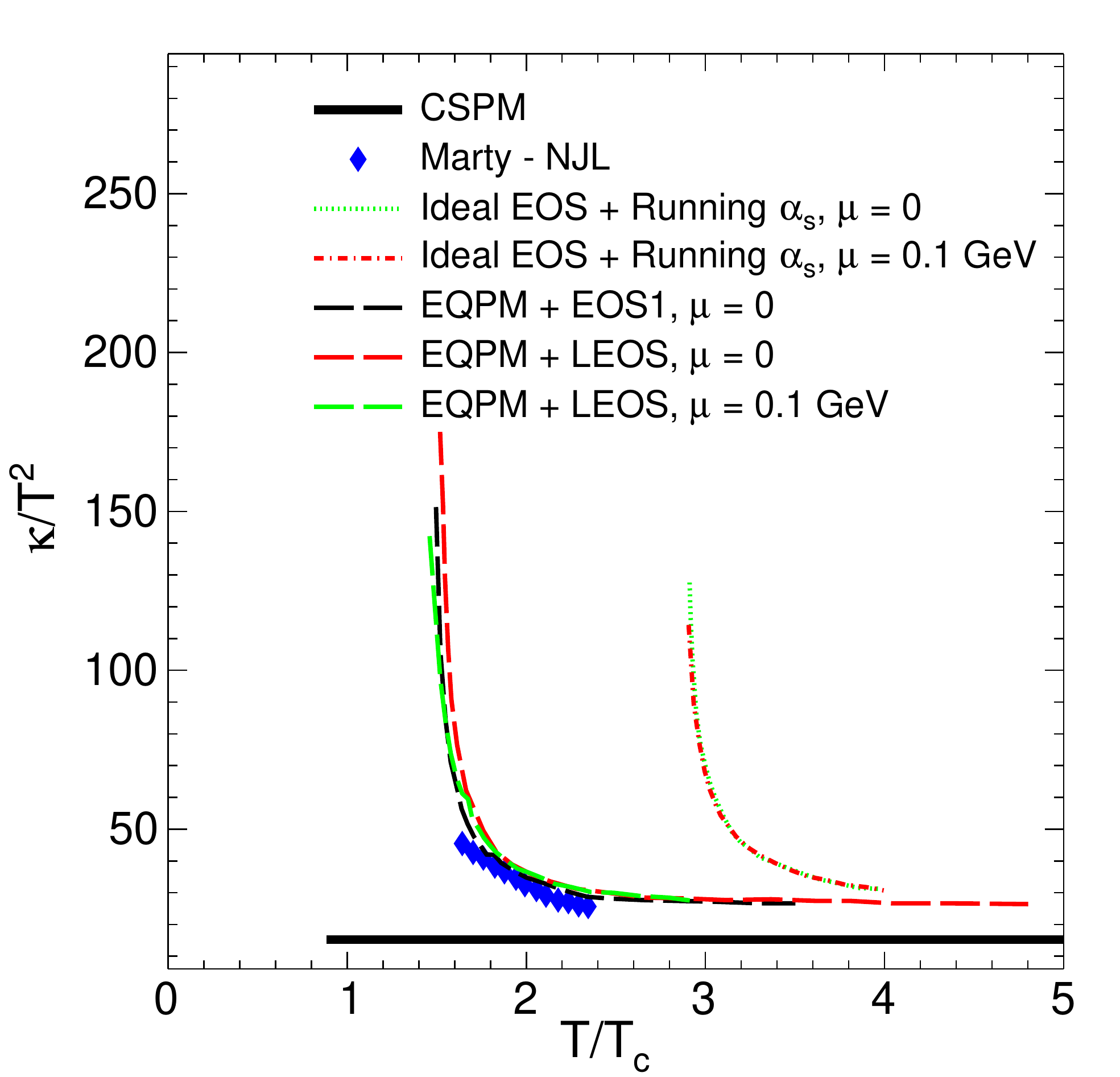}
\caption[]{(colour online) The dimensionless ratio of thermal conductivity to squared temperature ($\kappa/\rm T^{2}$) as a function of $\rm T/T_{c}$. The black solid line shows the results obtained in CSPM. The blue diamond symbols are the calculations of NJL model~\cite{Marty:2013ita}. The green dotted and the red dash-dotted lines represent the results of Ideal EOS with different running coupling constants. The black, red and green dashed lines are the results from a quasiparticle model.}
\label{kt2}
\end{figure}

\section{Results and Discussions}
\label{RD}
In this section, we discuss the results obtained in CSPM along with other theoretical results calculated in various approaches. We first time use color string percolation model to calculate thermal conductivity for hot QCD matter. For this, we use a very simple Drude model for electron. In CSPM, we assume that the strings break into equal number of quarks and antiquarks. 

The dimensionless quantity, $\kappa/T^{2}$ as a function of $\rm T/T_{c}$ is depicted in fig.~\ref{kt2}. The figure comprises CSPM result with various equation of states (EOSs) for both zero and nonzero quark chemical potentials. Different EOSs show different effects especially at lower temperatures and overlap with the results obtained in Ideal EOS. Our results from CSPM show almost independent behaviour with respect to $\rm T/T_{c}$ and are similar to almost all the results at higher temperatures, qualitatively. The blue diamond symbols are the results obtained within the ambit of Nambu-Jona-Lasinio (NJL) model for three quark flavors which decrease with temperature~\cite{Marty:2013ita}. The calculations of $\kappa$ obtained in Ideal EOS for QCD medium are also shown in the figure where non-interacting quarks and gluons are considered~\cite{Mitra:2017sjo}. The green dotted line and red dash-dotted line are the results for $\mu_q$ = 0 and 0.1 GeV, respectively. We have also shown the results obtained in effective fugacity quasiparticle model (EQPM)~\cite{Mitra:2017sjo}. The EQPM includes the hot QCD medium effects in terms of effective quasi-partons (quasi-gluons, quasi-quarks/anti-quarks). The hot QCD medium effects present in the hot QCD equations of states (EOSs), computed within either improved perturbative QCD (pQCD) or lattice QCD simulations are mapped onto the effective equilibrium distribution functions for the quasi-partons. In Ref.~\cite{Mitra:2017sjo}, the EQPM with QCD equation of state at $O(g^{5})$ (EOS1) and $O(g^{6} ln(1/g) + \delta)$ (EOS2), along with $2+1-$ flavor lattice QCD equation of state (LEOS) has been exploited. Results in all the cases shown in the figure, initially decrease rapidly with the temperature around twice the critical temperature and then saturates at a higher temperature. Our CSPM calculations show almost independent behaviour with respect to temperature. Our CSPM-based results for electrical conductivity in a Drude-like model have been reported recently \cite {Sahoo:2018dxn}.

\begin{figure}
\includegraphics[height=20em]{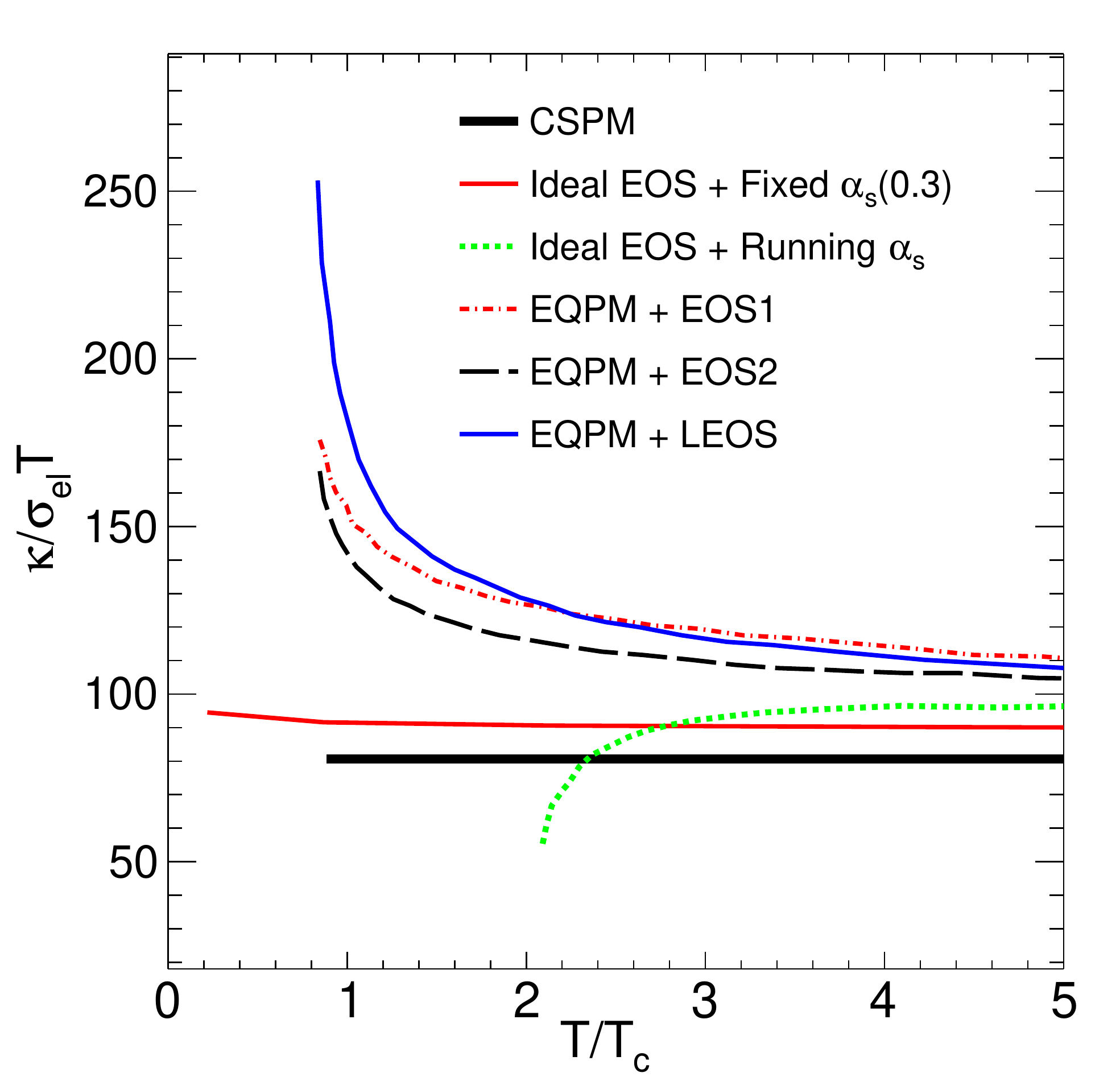}
\caption[]{(colour online) The results for Lorenz number versus $T/T_{c}$ are shown for various EOS. The black solid line shows the results calculated using CSPM. The red solid and green dotted lines represent the results of Ideal EOS for various coupling strength. The red dash-dotted, black dotted and blue lines are results obtained in different versions of quasiparticle model.}
\label{wfl}
\end{figure}

The interplay between electrical and thermal conductivity for any substance can be understood via studying the Wiedemann-Franz law which helps in understanding the relative importance between the charge diffusion and heat diffusion in any medim or substances. The ratio of thermal diffusion to charge diffusion or else called Lorenz number ($\mathbb{L}$) is depicted in Fig.~\ref{wfl}. For instance, the Lorenz number is constant for metals which infers that metals are good thermal conductors as well as electrical conductors. The study of the Lorenz number for hot QCD medium can throw light on the relation between electrical and thermal conduction in this medium. We find that, $\mathbb{L}$ is almost independent of $\rm T/T_{c}$ in CSPM. The value from CSPM is closer to the Stefan-Boltzmann (SB) limit of an ultra-relativistic gas of gluons and quarks. The solid red and green dashed lines are the results obtained in the famework of ideal EOS for quarks and gluons. The former is for the fixed strong coupling strength ($\alpha_s$ = 0.3), which shows a weak dependence on temperature particularly at a lower T, while the latter which is for running coupling constant first increases with T and becomes constant at a higher temperature. Again, we show the effective fugacity quasiparticle model (EQPM) results with various version of its equation of states~\cite{Mitra:2017sjo}. All the results obtained in EQPM shown by the solid blue, red dash-dotted and black dashed lines in the figure initially decrease rapidly and saturate at a higher T.      

\section{Summary and Outlook}
\label{summary}
In this work, for the first time we use a color string percolation approach to study the temperature dependence of thermal conductivity of hot QCD matter. We also study the ratio of thermal conductivity and electrical conductivity as a function of temperature. We have revisited the well known Wiedemann-Franz law for strongly interacting matter using CSPM. We compare our CSPM results with various versions of effective fugacity quasiparticle model and ideal equation of state for quarks and gluons degrees of freedom. We find a temperature independent thermal conductivity in CSPM while in other theoretical calculations thermal conductivity decreases upto T$\sim$1-2 T$_c$ and then saturates at a higher temperature. The ratio of thermal conductivity to electrical conductivity also shows a temperature independence behaviour CSPM approach and ideal EOS. EQPM results again show initial decrease with T and saturation afterwards. Since, the lattice QCD results of thermal conductivity are not available, a theoretical study of Wiedemann-Franz law for a hot QCD medium with color degrees of freedom could be important in understanding its behaviour near critical temperature. The Lorenz number for a hot QCD medium shows a similar behaviour like free electrons in  metals approaching to a number expected for a Stefan-Boltzmann limit of an ideal gas.

\section*{ACKNOWLEDGEMENTS}
PS and RNS acknowledge the financial supports from ALICE Project No. SR/MF/PS-01/2014-IITI(G) of Department of Science \& Technology, Government of India. SKT acknowledges the financial support from TEQIP-III, a joint venture of MHRD and the World Bank. RS gratefully acknowledges fruitful discussions with Pankaj R. Sagdeo.


\begin{thebibliography}{99}

\bibitem{STAR}
  B.~I.~Abelev {\it et al.} [STAR Collaboration],
  Phys.\ Rev.\ C {\bf 77},  054901 (2008).
  

  \bibitem{Luzum}
 M. Luzum and P. Romatschke,
 Phys. Rev. {\bf C 78}, 034915 (2008).
 
 
\bibitem{ALICE1}
  J.~Adam {\it et al.} [ALICE Collaboration],
  Phys.\ Rev.\ Lett.\  {\bf 117}, 182301 (2016).
  
\bibitem{ALICE2}
  J.~Adam {\it et al.} [ALICE Collaboration],
  Phys.\ Rev.\ Lett.\  {\bf 116}, 132302 (2016).

\bibitem{ALICE3}
  B.~Abelev {\it et al.} [ALICE Collaboration],
  Phys.\ Rev.\ Lett.\  {\bf 111}, 232302 (2013).
  
  
\bibitem{ALICE4}
  B.~Abelev {\it et al.} [ALICE Collaboration],
  Phys.\ Rev.\ Lett.\  {\bf 110},  012301 (2013).
  
  
\bibitem{ALICE5}
  J.~Adam {\it et al.} [ALICE Collaboration],
  Phys.\ Rev.\ C {\bf 93},  044903 (2016).

\bibitem{ALICE-JHEP}
  J.~Adam {\it et al.} [ALICE Collaboration],
  JHEP {\bf 1609}, 164 (2016).

   \bibitem{Mitra:2017sjo} 
  S.~Mitra and V.~Chandra,
  Phys.\ Rev.\ D {\bf 96}, 094003 (2017).


 \bibitem{Kapusta}
  J.~I.~Kapusta and J.~M.~Torres-Rincon,
  Phys.\ Rev.\ C {\bf 86},  054911 (2012).

\bibitem{Mitra:2018akk} 
  S.~Mitra and V.~Chandra,
  Phys.\ Rev.\ D {\bf 97}, 034032 (2018).
  
\bibitem{Armesto:1996kt} 
  N.~Armesto, M.~A.~Braun, E.~G.~Ferreiro and C.~Pajares,
  Phys.\ Rev.\ Lett.\  {\bf 77}, 3736 (1996).
  
  
  \bibitem{Nardi:1998qb} 
  M.~Nardi and H.~Satz,
  Phys.\ Lett.\ B {\bf 442}, 14 (1998).
  
  \bibitem{Braun:1999hv} 
  M.~A.~Braun and C.~Pajares,
  Eur.\ Phys.\ J.\ C {\bf 16}, 349 (2000).
  
  
  \bibitem{Braun:1997ch} 
  M.~A.~Braun, C.~Pajares and J.~Ranft,
  Int.\ J.\ Mod.\ Phys.\ A {\bf 14}, 2689 (1999).
  
  
  \bibitem{Braun:2000hd} 
  M.~A.~Braun and C.~Pajares,
  Phys.\ Rev.\ Lett.\  {\bf 85}, 4864 (2000).

\bibitem{Phyreport}
  M.A.~Braun {\it et al.} 
 Phys.\ Reports.\ {\bf 509}, 1 (2015).
  

   \bibitem{PLB642}
  J.~Dias de Deus and C.~Pajares,
  Phys.\ Lett.\ B {\bf 642}, 455 (2006).
  
 \bibitem{Gutay:2015cba} 
  L.~J.~Gutay {\it et al.},
  Int.\ J.\ Mod.\ Phys.\ E {\bf 24}, 1550101 (2015).
  

 \bibitem{Hirsch:2018pqm} 
  A.~S.~Hirsch, C.~Pajares, R.~P.~Scharenberg and B.~K.~Srivastava,
  arXiv:1803.02301 [hep-ph].


 \bibitem{Sahoo:2018dxn} 
  P.~Sahoo, S.~K.~Tiwari and R.~Sahoo,
  Phys.\ Rev.\ D {\bf 98}, 054005 (2018).


  \bibitem{Sahoo:2017umy} 
  P.~Sahoo, S.~K.~Tiwari, S.~De, R.~Sahoo, R.~P.~Scharenberg and B.~K.~Srivastava,
  Mod.\ Phys.\ Lett.\ A {\bf 34}, 1950034 (2019).


  \bibitem{Sahoo:2018dcz} 
  P.~Sahoo, S.~De, S.~K.~Tiwari and R.~Sahoo,
  Eur.\ Phys.\ J.\ A {\bf 54}, 136 (2018).



   \bibitem{Anderson}
  J. Anderson and H. Witting, Physica (Utrecht) {\bf 74}, 466 (1974).
  
  
  \bibitem{Greif:2014oia} 
  M.~Greif, I.~Bouras, C.~Greiner and Z.~Xu,
  Phys.\ Rev.\ D {\bf 90}, 094014 (2014).


  \bibitem{Israel:1979wp} 
  W.~Israel and J.~M.~Stewart,
  Annals Phys.\  {\bf 118}, 341 (1979).

  
\bibitem{groot}
  S. de Groot, W. van Leeuwen, and C. van Weert, RelativisticKinetic Theory: Principles and Applications (North-Holland,Amsterdam, 1980).


  \bibitem{Denicol:2012cn} 
  G.~S.~Denicol, H.~Niemi, E.~Molnar and D.~H.~Rischke,
  Phys.\ Rev.\ D {\bf 85}, 114047 (2012)
  Erratum: [Phys.\ Rev.\ D {\bf 91}, 039902 (2015)].


  \bibitem{Greif:2013bb} 
  M.~Greif, F.~Reining, I.~Bouras, G.~S.~Denicol, Z.~Xu and C.~Greiner,
  Phys.\ Rev.\ E {\bf 87}, 033019 (2013).
  
\bibitem{link}
  $https://unlcms.unl.edu/cas/physics/tsymbal/teaching/SSP-927/Section\%2008_Electron_Transport.pdf$

  \bibitem{Heiselberg:1993cr} 
  H.~Heiselberg and C.~J.~Pethick,
  Phys.\ Rev.\ D {\bf 48}, 2916 (1993).

\bibitem{link1}
  $https://unlcms.unl.edu/cas/physics/tsymbal/teaching/SSP-927/Section\%2007_Free_Electron_Model.pdf$


  \bibitem{Marty:2013ita} 
  R.~Marty, E.~Bratkovskaya, W.~Cassing, J.~Aichelin and H.~Berrehrah,
  Phys.\ Rev.\ C {\bf 88}, 045204 (2013).
  
  

   
\end{thebibliography}
\end{document}